\documentclass[conference]{IEEEtran}

\usepackage{cite}
\usepackage{amsmath}
\usepackage{amsthm}
\usepackage{url}

\theoremstyle{definition}
\newtheorem{protocol}{Protocol}

\begin{document}

\title{A New Key Establishment Protocol and its Application in Pay-TV Systems}
\author{
\IEEEauthorblockN{Peter Roelse}
\IEEEauthorblockA{Advanced Development \& Innovation\\
Irdeto B.V.\\
Hoofddorp, The Netherlands\\
peter.roelse@irdeto.com}
}

\maketitle

\begin{abstract}
A pay-TV consumer uses a decoder to access encrypted digital content. To this end, the decoder contains a chip capable of decrypting the content if provisioned with the appropriate content decryption keys. A key establishment protocol is used to secure the delivery of the content decryption keys to the chip. This paper presents a new protocol and shows how the protocol can be applied in a pay-TV system. Compared to existing protocols, the presented solution reduces the risk that decoders need to be replaced in order to correct a security breach. The new protocol has recently been incorporated in an ETSI standard.
\end{abstract}


\IEEEpeerreviewmaketitle

\section{Introduction}
In a pay-TV system, the pay-TV provider's head-end system encrypts the content before broadcasting it, and a consumer uses a decoder capable of decrypting the content to access it. In order to achieve this, a decoder contains a chip that implements the content decryption algorithm; this chip is also referred to as the content decryption chip in this paper. Examples of a decoder are a set-top box or a PC Card in case of a Common Interface (CI) or CI+ module.

A pay-TV provider uses a Conditional Access (CA) system to control access to the content, thereby ensuring that only authorized decoders have access to the keys required to decrypt the content. In particular, each decoder associated with the pay-TV provider contains a CA client, and only the CA client of an authorized decoder will pass content decryption keys, referred to as control words in a pay-TV system, to the content decryption chip in the decoder.

Attacks in which an adversary compromises and re-distributes control words are a threat to the security of a pay-TV system, as this enables non-authorized access to the corresponding content. It is generally easy for an adversary to read or modify messages passed from the CA client to the content decryption chip in a decoder, or to inject messages into this channel. For these reasons, a cryptographic protocol is used to transport control words from the CA client to the content decryption chip.

This paper presents a new key establishment protocol. The new protocol offers a similar level of security as existing protocols (see e.g.\ \cite{KLAD10}) against attacks in which content decryption keys are compromised and re-distributed. The main advantage of the new protocol is that it achieves the unique and desirable property of being able to restore security for future protocol executions without the need to replace any decoder in the event that all system components other than the content decryption chips have been compromised. By comparison, existing protocols necessitate the replacement of the entire decoder population in this scenario. The new protocol was recently incorporated in a new ETSI standard \cite{ECI17}.

\section{Preliminaries}
\subsection{Basic Concepts of a Pay-TV System}
This section describes aspects of a pay-TV system that are relevant to this paper. This paper assumes that the pay-TV system is compliant with the widely adopted Digital Video Broadcasting (DVB) standard (see also \url{www.dvb.org}).

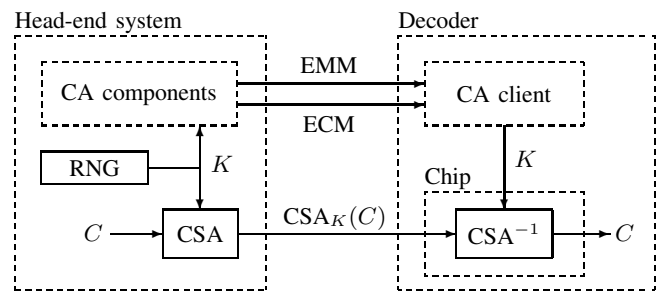
\begin{figure}[b]
\centering
\small
\begin{picture}(242,107)(0,0)
\put(0,0){\dashbox{2}(95,96)}
\put(0,99){Head-end system}
\put(10,40){\framebox(40,12){RNG}}
\put(50,46){\line(1,0){20}}
\put(70,46){\vector(0,1){16}}
\put(70,46){\vector(0,-1){16}}
\put(74,44.5){$K$}
\put(10,62){\dashbox{2}(74,24){CA components}}
\put(56,12){\framebox(28,18){CSA}}
\put(36,21){\vector(1,0){20}}
\put(26,18.5){$C$}
\put(84,78){\vector(1,0){71}}
\put(108,82){EMM}
\put(84,70){\vector(1,0){71}}
\put(109,59.5){ECM}
\put(84,21){\vector(1,0){83}}
\put(101.5,25){CSA$_K(C)$}
\put(145,0){\dashbox{2}(97,96)}
\put(145,99){Decoder}
\put(155,62){\dashbox{2}(60,24){CA client}}
\put(185,62){\vector(0,-1){32}}
\put(189,46.5){$K$}
\put(155,5){\dashbox{2}(60,32)}
\put(155,40){Chip}
\put(167,12){\framebox(36,18){CSA$^{-1}$}}
\put(203,21){\vector(1,0){22}}
\put(227,18.5){$C$}
\end{picture}
\caption{DVB pay-TV system.}
 \label{fig:1}
\end{figure}

Fig.~\ref{fig:1} depicts the basic components of a DVB pay-TV system. The pay-TV provider operates a head-end system and a consumer uses a decoder to access content broadcast by the provider. Note that only one decoder is depicted in Fig.~\ref{fig:1}; in practice a large number of decoders are associated with the head-end system. To protect content, denoted by $C$ in Fig.~\ref{fig:1}, it is encrypted inside the head-end system before it is broadcast. To this end, DVB defines a symmetric encryption scheme referred to as the Common Scrambling Algorithm (CSA) \cite{CSA3}. A CSA key is referred to as a control word and denoted by $K$ in Fig.~\ref{fig:1}. Throughout the paper, the key of a keyed cryptographic algorithm is written as a subscript; for example, the encryption of $C$ using the CSA encryption algorithm and control word $K$ is denoted by CSA$_K(C)$. The corresponding decryption algorithm CSA$^{-1}$ is implemented inside the content decryption chip integrated in the decoder. Typically, a control word is updated every 5 to 10 seconds.

A CA system ensures that only authorized decoders can access content. As shown in Fig.~\ref{fig:1}, the CA system comprises CA components at the head-end and a CA client in each decoder. The DVB standard does not specify the CA system itself. Consequently, a number of different CA system suppliers exist, each offering their own proprietary CA system. However, DVB has defined a head-end system architecture, referred to as the DVB SimulCrypt standard \cite{DVB1} \cite{DVB2}. This standard identifies the logical components in the head-end system and it specifies the interfaces between these components. In particular, in the SimulCrypt standard the head-end system implements the CSA encryption algorithm and a Random Number Generator (RNG) to generate control words, which are supplied as input to CSA and to the CA components at the head-end (see also Fig.~\ref{fig:1}). An important property of SimulCrypt is that this architecture enables the use of multiple CA systems in the head-end system, each protecting the same encrypted content (in Fig.~\ref{fig:1} only one CA system is depicted). Such systems are referred to as interoperating CA systems.

Further, DVB defines two types of CA message that can be sent from the CA components to a CA client \cite{CSA3}. The first type is an Entitlement Management Message (EMM); an EMM is typically used to authorize a decoder to access a specific piece of content. The second type is an Entitlement Control Message (ECM); ECMs are used to distribute control words to the CA clients of authorized decoders. The contents of EMMs and ECMs are proprietary to the CA system.

The CA components at the head-end generate the CA messages before broadcasting the encrypted content and the CA messages to the decoders. An important property for the protocol described in this paper is that an electronic return channel from a decoder to the head-end system may not be available in a pay-TV system, in particular if a satellite or terrestrial network is used for broadcasting information to the decoders. This paper therefore assumes that such a return channel is not available.

There is usually a limited amount of bandwidth available for sending CA messages in a pay-TV system, as a pay-TV provider prefers to use as much of the available bandwidth as possible for broadcasting content. In particular, it is not possible to distribute a uniquely encrypted message to every CA client for every control word $K$. To address this, the CA system can protect a number of control words using a key that is shared between all the CA clients of authorized decoders.

The CA client in a decoder processes EMMs and ECMs. If the decoder is authorized to access content associated with a specific ECM, that is, if the CA client in the decoder has received and processed an EMM authorizing the decoder to access this content, then the CA client derives the control word from the ECM. Next, the CA client passes the control word to the content decryption chip in the decoder. It is generally possible to update the CA client in a decoder. In particular, this enables the pay-TV provider to correct any security breach of the CA clients without replacing any decoder.

\subsection{Control Word Re-distribution Attacks}
In a control word re-distribution attack an adversary first compromises control words, e.g.\ by extracting control words from an authorized decoder. Next, the adversary re-distributes the control words to pirate decoders that have access to the pay-TV provider's broadcast. A pirate decoder then uses the encrypted content in the broadcast and the re-distributed control words as inputs to its implementation of CSA$^{-1}$ to illegally access content in real-time.

The content decryption algorithm CSA$^{-1}$ may only be implemented after obtaining a license from ETSI. This makes it possible to suppress illegal implementations. Furthermore, CSA is a DVB-confidential cipher and CSA v3 contains an emulation resistant algorithm (see also \cite{CSA3}). These measures make it difficult for an adversary to use pirate decoders containing illegal implementations of CSA$^{-1}$ in a control word re-distribution attack. However, it is generally easy to access the channel from the CA client to the content decryption chip in a decoder containing a legitimate implementation of CSA$^{-1}$ (see also Fig.~\ref{fig:1}). In particular, if this channel is unprotected, then an adversary may compromise control words when they are passed from the CA client to the chip, or the adversary may inject compromised control words into this channel, using the decoder as a pirate decoder. In practice, the channel from the CA client to the content decryption chip is therefore protected using a cryptographic protocol. Such a protocol should provide implicit key authentication so that only the content decryption chips of authorized decoders have access to the control word. As explained later in Section~\ref{sec:pay-TV}, this measure is also useful for identifying compromised chips that are used as source devices in a control word re-distribution attack. In addition, the protocol should prevent the adversary from finding protocol messages that enable the content decryption chip of a non-authorized decoder that is compliant with the protocol to derive a compromised control word. This is referred to as protecting the authenticity of protocol messages and this measure makes it difficult for an adversary to use a compliant chip containing a legitimate CSA$^{-1}$ implementation as a sink device in a control word re-distribution attack in the case that the values of the control words are known.

\section{The New Key Establishment Protocol}
\subsection{Description of the Protocol}
\label{sec:description}
This section presents the new key establishment protocol, and the next section shows how the protocol can be applied in a pay-TV system. The protocol takes into account that there is generally no trust relation between pay-TV providers, and that there is a limited amount of bandwidth available for sending CA messages. The setting in this paper is that multiple receivers are associated with one sender. The protocol enables the sender and a number of receivers selected by the sender to derive a shared secret $K$. A receiver selected by the sender is referred to as an authorized receiver in the following text. The protocol provides implicit key authentication; in other words, the sender is assured that only authorized receivers have access to $K$. In addition, the protocol protects the authenticity of protocol messages in that it is computationally infeasible for an adversary to find protocol messages that enable a non-authorized receiver that is compliant with the protocol to derive a given $K$ (as generated by the sender).

\begin{figure*}[!t]
\centering
\small
\begin{picture}(420,166)(0,0)
\put(0,0){\dashbox{2}(150,156)}
\put(0,158){$A$}
\put(10,130){\framebox(40,12){KPG}}
\put(50,132){\line(1,0){80}}
\put(130,132){\vector(0,-1){52}}
\put(107,107){$SK_A$}
\qbezier(10,112)(30,117)(50,112)
\qbezier(10,112)(30,107)(50,112)
\put(10,112){\line(0,-1){20}}
\put(50,112){\line(0,-1){20}}
\qbezier(10,92)(30,87)(50,92)
\put(20,96){$PK_B$}
\put(50,102){\line(1,0){40}}
\put(70,107){$PK_B$}
\put(90,102){\vector(0,-1){22}}
\put(10,64){\framebox(40,12){RNG}}
\put(50,70){\vector(1,0){30}}
\put(55,75){$LK_B$}
\put(70,40){\line(1,0){20}}
\put(70,40){\line(0,1){30}}
\put(80,60){\framebox(20,20){$E$}}
\put(100,70){\vector(1,0){20}}
\put(120,60){\framebox(20,20){$S$}}
\put(10,14){\framebox(40,12){RNG}}
\put(50,20){\vector(1,0){30}}
\put(62,25){$r$}
\put(80,10){\framebox(20,20){$e$}}
\put(90,40){\vector(0,-1){10}}
\put(50,140){\vector(1,0){320}}
\put(204.5,127){$PK_A$}
\put(140,70){\vector(1,0){150}}
\put(168,77){$S_{SK_A}(B, E_{PK_B}(LK_B))$}
\put(100,20){\vector(1,0){230}}
\put(198,27){$e_{LK_B}(r)$}
\put(280,0){\dashbox{2}(140,156)}
\put(280,158){$B$}
\put(290,60){\framebox(20,20){$V$}}
\put(300,140){\vector(0,-1){60}}
\put(310,70){\vector(1,0){20}}
\put(330,60){\framebox(20,20){$D$}}
\put(340,90){\vector(0,-1){10}}
\put(332,93){$SK_B$}
\put(340,60){\vector(0,-1){30}}
\put(317,42){$LK_B$}
\put(330,10){\framebox(20,20){$d$}}
\put(350,20){\vector(1,0){20}}
\put(357,25){$r$}
\put(370,10){\framebox(20,136){$h$}}
\put(390,78){\vector(1,0){15}}
\put(407,75.5){$K$}
\end{picture}
\caption{The new key establishment protocol.}
\label{fig:2}
\end{figure*}
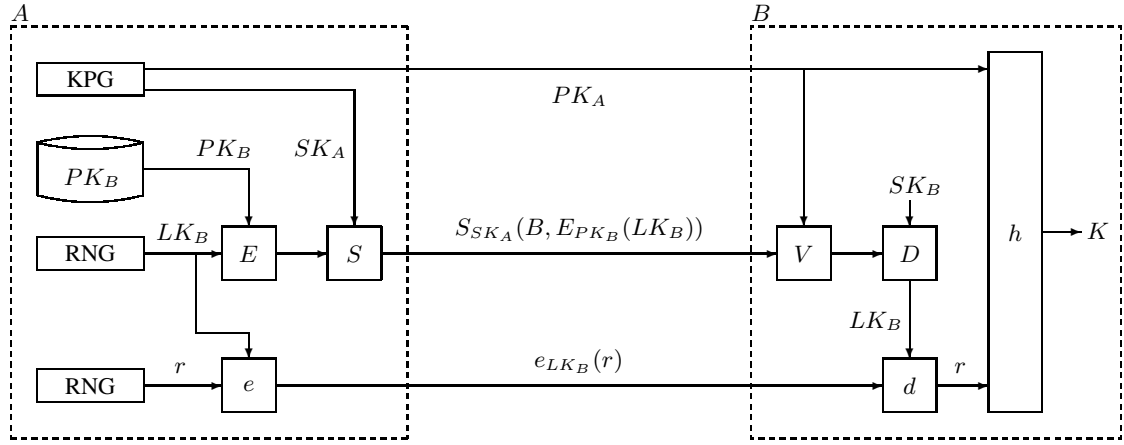

The new protocol is described in Protocol~\ref{protocol}, and the parties and their protocol steps are depicted in Fig.~\ref{fig:2} (with the exception of Step~\ref{HE6}). The protocol uses a public-key encryption scheme and a digital signature scheme. The public-key encryption and decryption algorithms are denoted by $E$ and $D$, respectively. Further, the signature generation algorithm and the signature verification algorithm are denoted by $S$ and $V$. For ease of notation, a signature scheme with message recovery is used in the description of the protocol. Instead, a signature scheme with partial message recovery or a signature scheme with appendix can be used. The identity of receiver $B$ is denoted by $B$ in signed messages, and in Fig.~\ref{fig:2}, KPG denotes a key pair generator that generates key pairs associated with the signature scheme. The protocol also uses a symmetric encryption scheme and a cryptographic hash function $h$. The corresponding encryption and decryption algorithms are denoted by $e$ and $d$, respectively. Further, it is assumed that the bit-length of the output of $h$ equals the bit-length of $K$. A truncation method may be applied to the output of a well-known hash function (e.g.\ SHA-512) in order to define a suitable $h$. A trusted third party acts as a certification authority in the protocol. The following description therefore assumes that the trusted third party has generated a key pair associated with the signature scheme. The private key and the public key in this pair are denoted by $SK_T$ and $PK_T$.

\begin{protocol}\label{protocol} {\ } \\
\textbf{Initialization:} each receiver generates a key pair associated with the public-key encryption scheme. The private key and the public key in the key pair of receiver $B$ are denoted by $SK_B$ and $PK_B$, respectively. Initialize $B$ with $SK_B$, and securely transfer $PK_B$ to trusted third party $T$.\\ \vspace{-3mm} \\
\textbf{I. Establish long-term key:}
\begin{enumerate}
\item \label{HE1} $A$ generates a key pair associated with the signature scheme. The private key and the public key in this pair are denoted by $SK_A$ and $PK_A$, respectively.
\item For each receiver $B$ associated with $A$:
\begin{enumerate}
\item \label{HE2} $A$ generates a symmetric long-term key $LK_B$.
\item $A$ computes $E_{PK_B}(LK_B)$ (it is assumed that $A$ has received $PK_B$ from $T$; $PK_B$ is stored in a database in Fig.~\ref{fig:2}).
\item \label{HE4} $A$ computes the signature\\ $S_{SK_A}(B, E_{PK_B}(LK_B))$.
\item \label{CC1} $A$ sends to $B$ the values $PK_A$ and\\ $S_{SK_A}(B, E_{PK_B}(LK_B))$.
\item $B$ computes $V_{PK_A}(S_{SK_A}(B, E_{PK_B}(LK_B)))$. If the signature is valid, then $B$ recovers $(B, E_{PK_B}(LK_B))$ from the signature, and verifies if $B$ is the intended recipient. $B$ aborts the protocol if any check fails.
\item $B$ computes $LK_B = D_{SK_B}(E_{PK_B}(LK_B))$.
\end{enumerate}
\end{enumerate}
\textbf{II. Establish shared secret:}
\begin{enumerate}
\setcounter{enumi}{2}
\item \label{HE5} $A$ generates a secret random number $r$ with a bit-length that is equal to, or slightly larger than, the bit-length of $K$.
\item \label{HE6} $A$ represents $PK_A$ and $r$ by fixed-length bit strings, concatenates these bit strings to obtain $r || PK_A$ and computes $K = h(r || PK_A)$.
\item For each authorized receiver $B$:
\begin{enumerate}
\item \label{CC2} $A$ computes $e_{LK_B}(r)$.
\item \label{CC3} $A$ sends to $B$ the value $e_{LK_B}(r)$.
\item $B$ computes $r = d_{LK_B}(e_{LK_B}(r))$.
\item $B$ computes the shared secret $K = h(r || PK_A)$ (the concatenation is not depicted in Fig.~\ref{fig:2}).
\end{enumerate}
\end{enumerate}
\end{protocol}

The initialization phase is a one-time setup of the protocol. Observe that the distribution of $PK_B$ from $T$ to $A$ is not described in Protocol~\ref{protocol}. In practice, $T$ can first securely transfer $PK_T$ to $A$. Next, $T$ can create the certificate $S_{SK_T}(B, PK_B)$ and send this certificate to $A$. $T$ can also manage the revocation of such certificates. Based on this information, $A$ can create a database containing public keys of non-revoked certificates (as depicted in Fig.~\ref{fig:2}).

The remainder of the protocol is divided into two similar phases: in Phase I the sender transports a long-term key to each receiver, and in Phase II the sender transports a shared secret to each authorized receiver. In practice, Phase II is executed a number of times after Phase I is executed, and the sender can select any set of authorized receivers for every execution of Phase II. The protocol description assumes that there is only one sender; however, in practice multiple senders can independently execute the protocol with receiver $B$ based on the same initialization value $SK_B$.

\subsection{Security Analysis}
\label{sec:security}
For ease of exposition, the channels between a receiver and the trusted third party (as used during the initialization of the protocol) and between a sender and the trusted third party are assumed to be secure. The remainder of this section assumes that $B'$ is a receiver that is: (1) compliant with Protocol~\ref{protocol}, (2) non-authorized in the execution of Phase II in which $K$ is generated and (3) authorized in all other executions of Phase II. Further, it is assumed that all other receivers are authorized in all executions of Phase II. These assumptions ensure that $B'$ and the adversary have as much information as possible in the discussions below.

To show that only authorized receivers have access to the shared secret $K$, it is assumed that the following information is available to $B'$ (in addition to publicly known information): (1) the values of all secret random numbers except $r$ (implying that $B'$ also has access to all the corresponding shared secrets), (2) the value of $SK_{B'}$, and (3) all values of $LK_{B'}$. Note that the random number $r$ corresponds to exactly one execution of Phase I. The ciphertexts, random numbers, shared secrets and values of $LK_{B'}$ which are available to $B'$ and which are associated with other executions of Phase I (and possibly with other senders) do not make it easier for $B'$ to find $r$ since anyone can easily generate link keys, random numbers, shared secrets, and ciphertexts that are indistinguishable from the ones generated by the legitimate senders based on publicly known information. This means that the problem of finding $r$ is equivalent to finding $r$ from the available data that are associated with the particular execution of Phase I as mentioned above and the corresponding executions of Phase II. From the observations that the private keys $SK_{B}$ and the long-term keys $LK_{B}$ are generated independently and at random for each receiver $B$ and that a fresh random number is generated at random in every execution of Phase II, and under the assumption that the encryption algorithms do not leak information about their key and message inputs, it follows that $B'$ has no access to the random number $r$. From this, and under the assumption that $h$ behaves as a random function, it follows that $B'$ does not have access to the shared secret $K$.

With respect to the protection of the authenticity of protocol messages, suppose that an adversary manages to find three protocol messages $m_1$, $m_2$ and $m_3$ that enable a non-authorized receiver $B'$ that is compliant with Protocol~\ref{protocol} to derive a given shared secret $K$ (as generated by $A$). The inputs to $V$ and $d$ (see also Fig.~\ref{fig:2}) are referred to as $m_2$ and $m_3$ respectively, and the third message is referred to as $m_1$. The following discussion assumes that the adversary has access to the values of all secret random numbers (and, consequently, to all the shared secrets). Under the assumption that $h$ is second preimage resistant, the input to $h$ must be $r || PK_A$. This implies that $m_1 = PK_A$. The adversary now needs to find $m_2$ and $m_3$ such that the input to $D$ equals $E_{PK_{B'}}(LK)$ for some $LK$ and such that $d_{LK}(m_3) = r$. Furthermore, $m_2$ needs to be accepted by the signature verification algorithm using $PK_A$ as input. Under the assumption that the signature scheme is secure against signature forgery attacks, the only option for the adversary is to use the protocol message $S_{SK_A}(B', E_{PK_{B'}}(LK_{B'}))$ as generated by $A$ in Step~\ref{HE4} and distributed to $B'$ in Step~\ref{CC1} in the execution of Phase I associated with $(SK_A, PK_A)$ as $m_2$. This implies that $m_3 = e_{LK_{B'}}(r)$. A similar reasoning as that for $r$ in the previous paragraph can be applied to show that the problem of finding $m_3$ is equivalent to finding $m_3$ from the available data that are associated with this execution of Phase I and the corresponding executions of Phase II. If one or more ciphertexts were distributed to $B'$ in these executions of Phase II, then let these ciphertexts be denoted by $e_{LK_{B'}}(r_i)$ for $1 \leq i \leq m$ with $m \geq 1$. Since the keys $SK_{B}$, $LK_{B}$ and the shared secrets are all generated independently and at random, and under the assumptions that the encryption algorithms do not leak information about $LK_{B'}$, it follows that for all practical values of $m$ the only useful information for the adversary is that $m_3 \neq e_{LK_{B'}}(r_i)$ for $1 \leq i \leq m$. If no ciphertexts were distributed to $B'$ in these executions, then no useful information is available to the adversary. From this it follows that the adversary is not able to find $m_3$.

Multiple senders can execute the new protocol independently with $B$ based on the same initialization value $SK_B$ and without requiring a trust relation between senders since a compromise of all keys of a sender does not affect the operations of other senders. In particular, observe that the new protocol prevents that sender $A'$ can generate protocol messages that enable a receiver that is compliant with the protocol to derive a given shared secret $K$ (as generated by $A$) since the hash function $h$ binds the shared secret $K$ to a specific key pair $(SK_A, PK_A)$. The binding also enables $A$ to revoke a key pair $(SK_A, PK_A)$ and all corresponding long-term keys $LK_B$.

If an adversary compromises the system of a sender or the system of the trusted third party (i.e.\ compromising all secrets of such a system), then one or both security properties of the protocol may no longer be satisfied. An important aspect of the protocol when applied in a pay-TV system is whether security be restored for future protocol executions based on the same initialization values. The main difference with existing protocols that offer implicit key authentication and protection of the authenticity of protocol messages is that the trusted third party does not manage any secret associated with a value that was used to initialize $B$ in the new protocol. This results in the unique property of enabling the restoration of security for future protocol executions (based on $SK_B$) even in the event that the system of the trusted third party and the systems of the senders have been compromised, in other words, if $SK_T$ and all keys of the senders' systems have been compromised. To see this, observe that a compromise of $SK_T$ invalidates the certificate of $B$ (i.e.\ $S_{SK_T}(B, PK_B)$), but not the key pair $(SK_B, PK_B)$ or certificate revocation information. The trusted third party and the senders can first restore the security of their systems to correct the security breach. Next, the trusted third party can update their key pair $(SK_T, PK_T)$. After this, the trusted third party can securely transfer the public key of this updated key pair to the senders, and re-issue certificates using the private key of the updated pair. Finally, senders can set up their system as if they were new senders joining the system, using the re-issued certificates and newly generated keys and shared secrets in their future protocol executions. From the discussions above it follows that both security properties are restored for these executions.

\section{Application in Pay-TV Systems}
\label{sec:pay-TV}

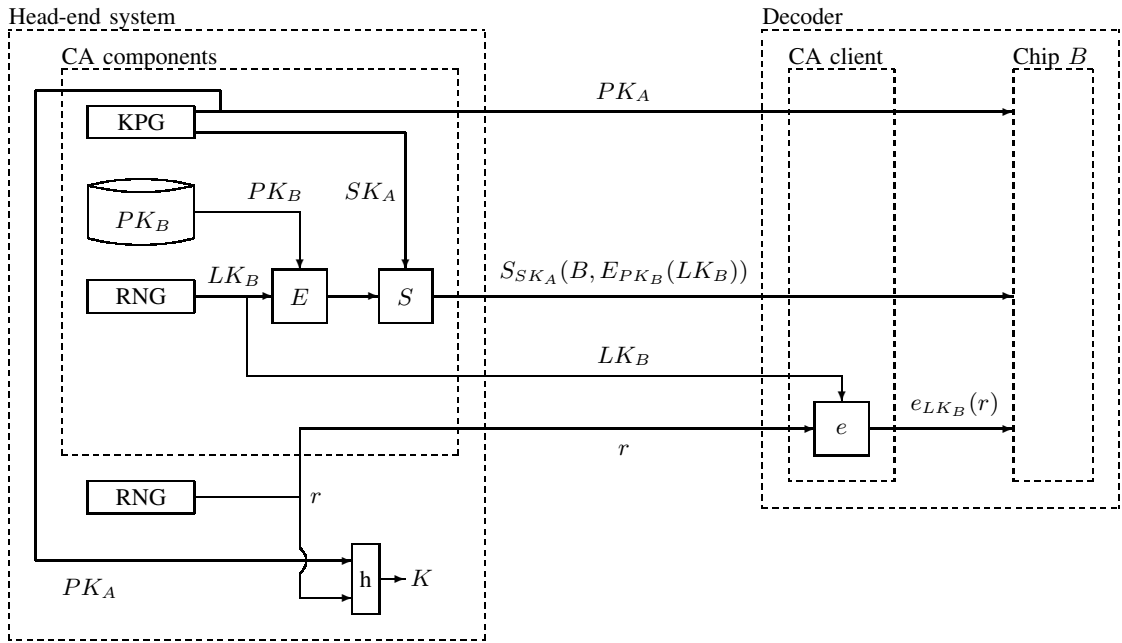
\begin{figure*}[!t]
\centering
\small
\begin{picture}(420,240)(0,0)
\put(0,0){\dashbox{2}(180,231)}
\put(0,233){Head-end system}
\put(30,48){\framebox(40,12){RNG}}
\put(70,54){\line(1,0){40}}
\put(110,35){\line(0,1){45}}
\qbezier(110,35)(115,30)(110,25)
\put(110,16){\line(0,1){9}}
\put(110,16){\vector(1,0){20}}
\put(114,52){$r$}
\put(130,10){\framebox(10,26){h}}
\put(140,23){\vector(1,0){10}}
\put(152,20.5){$K$}
\put(20,70){\dashbox{2}(150,146)}
\put(20,218){CA components}
\put(30,190){\framebox(40,12){KPG}}
\put(80,200){\line(0,1){8}}
\put(80,208){\line(-1,0){70}}
\put(10,208){\line(0,-1){178}}
\put(10,30){\vector(1,0){120}}
\put(20,17){$PK_A$}
\put(70,192){\line(1,0){80}}
\put(150,192){\vector(0,-1){52}}
\put(127,167){$SK_A$}
\qbezier(30,172)(50,177)(70,172)
\qbezier(30,172)(50,167)(70,172)
\put(30,172){\line(0,-1){20}}
\put(70,172){\line(0,-1){20}}
\qbezier(30,152)(50,147)(70,152)
\put(40,156){$PK_B$}
\put(70,162){\line(1,0){40}}
\put(110,162){\vector(0,-1){22}}
\put(90,167){$PK_B$}
\put(30,124){\framebox(40,12){RNG}}
\put(70,130){\vector(1,0){30}}
\put(75,135){$LK_B$}
\put(90,130){\line(0,-1){30}}
\put(100,120){\framebox(20,20){$E$}}
\put(120,130){\vector(1,0){20}}
\put(140,120){\framebox(20,20){$S$}}
\put(70,200){\vector(1,0){310}}
\put(222,205){$PK_A$}
\put(160,130){\vector(1,0){220}}
\put(185.5,137){$S_{SK_A}(B, E_{PK_B}(LK_B))$}
\put(90,100){\line(1,0){225}}
\put(222.5,105){$LK_B$}
\put(110,80){\vector(1,0){195}}
\put(230.5,70){$r$}
\put(285,50){\dashbox{2}(135,181)}
\put(285,233){Decoder}
\put(380,60){\dashbox{2}(30,156)}
\put(380,218){Chip $B$}
\put(295,60){\dashbox{2}(40,156)}
\put(295,218){CA client}
\put(305,70){\framebox(20,20){$e$}}
\put(315,100){\vector(0,-1){10}}
\put(325,80){\vector(1,0){55}}
\put(341,87){$e_{LK_B}(r)$}
\end{picture}
\caption{Application in a pay-TV system.}
\label{fig:3}
\end{figure*}

In a pay-TV system, sender $A$ comprises a CA system and components in the head-end system that are shared between interoperating CA systems. Further, receiver $B$ is a content decryption chip in a decoder associated with $A$, and $K$ is a control word. Fig.~\ref{fig:3} depicts an example of how the new protocol can be applied in a pay-TV system. As in DVB SimulCrypt, the example keeps the number of secret and private keys used outside the CA components in the head-end system as small as possible, while still facilitating interoperability. The following text describes which of the protocol steps are performed by the CA components at the head-end and the CA clients in the decoders, respectively. The communications from the CA components to the CA clients are also described in this section. Note that these aspects are not described in Protocol~\ref{protocol}, as the CA components and the CA clients are both part of the sender in Protocol~\ref{protocol}. In particular, the communications from the sender to the receiver as shown in Fig.~\ref{fig:2} are the communications from the CA client to the chip in Fig.~\ref{fig:3}.

As shown in Fig.~\ref{fig:3}, the random number generator and the hash function $h$ are the shared components; these components perform Steps~\ref{HE5} and \ref{HE6} of the protocol. As the public key $PK_A$ is input to $h$, the CA components output this value after generating the corresponding key pair in Step~\ref{HE1}. The public key $PK_A$ is also distributed from the CA components to the CA clients, as it needs to be distributed to the chips in Step~\ref{CC1}. An EMM can be used to distribute this value to all the CA clients. Recall that $PK_A$ does not need to be protected during its distribution to the chips. In particular, no certificates or certificate revocation information need to be distributed to a decoder in the application of the new protocol.

The signature $S_{SK_A}(B, E_{PK_B}(LK_B))$ is generated uniquely for chip $B$ in Step~\ref{HE4}. However, these values are only generated during the execution of Phase I, and not every 5 to 10 seconds in case of $e_{LK_B}(K)$ in Phase II. This makes it feasible to distribute the values $S_{SK_A}(B, E_{PK_B}(LK_B))$ from the head-end system to the CA clients. The CA components at the head-end can therefore perform Steps~\ref{HE2} --~\ref{HE4}. Next, the value $S_{SK_A}(B, E_{PK_B}(LK_B))$ is distributed to the CA client associated with $B$, along with a copy of $LK_B$ to enable the CA client to perform Step~\ref{CC2}. This distribution can be done using an EMM. After receiving the EMM, the CA client passes the value $S_{SK_A}(B, E_{PK_B}(LK_B))$ to $B$ in Step~\ref{CC1}.

The random number $r$ is input to the CA components at the head-end instead of the shared secret $K$ in Fig.~\ref{fig:1}. Next, the CA components include $r$ in an ECM instead of $K$ and distribute the ECM to the CA clients. The confidentiality and authenticity of $r$ can be protected with the same methods as used for protecting the confidentiality and authenticity of $K$ in a legacy CA system. After receiving the ECM, the CA clients of authorized decoders derive $r$ from the ECM. Next, the CA client associated with chip $B$ performs Step~\ref{CC2} and passes the value $e_{LK_B}(r)$ to $B$ in Step~\ref{CC3}.

If CA systems interoperate at the head-end, then each of these CA systems is associated with a sender as defined in Section~\ref{sec:description}. Each sender executes the new protocol independently with their decoders, with the exception of Steps~\ref{HE5} and \ref{HE6}; these steps are shared between all the interoperating senders and performed by the shared components at the head-end. The random number is provided as input to the CA components of each of the interoperating CA systems. The input to $h$ can be easily generalized so that it can contain all the public keys as output by the CA components, and a chip can accept a message that is signed with any of the corresponding private keys. Observe that this implies that the chip in a decoder also needs access to all these public keys. To achieve this, the public key of each CA system can be provided as input to the CA components of all other interoperating CA systems, which can then distribute the public keys to their decoders.

Recall from Section~\ref{sec:description} that multiple senders can execute the new protocol independently with any receiver without requiring a trust relation between senders. If the protocol is applied in a pay-TV system and if the CA systems associated with senders $A$ and $A'$ interoperate at the head-end, then $A$ and $A'$ do not execute the protocol independently with their receivers. In particular, $A$ and $A'$ share the secret $K$, implying that a trust relation is required between interoperating senders. However, observe that such senders are always part of the same head-end system. As a head-end system is assumed to be operated by a single pay-TV provider, any pay-TV provider can use a compliant decoder in their operation without the need to trust any other pay-TV provider.

As the protocol provides implicit key authentication, a consumer with a non-authorized decoder cannot obtain illegal access to content by compromising all secrets of their decoder's chip. Additionally, the re-distribution of a compromised $K$ to non-authorized and non-compromised chips that are compliant with the protocol is prevented since the authenticity of protocol messages is protected. From the discussion in Section~\ref{sec:security} it follows that both these properties can be restored for future protocol executions without replacing any decoder in the event that the system of the trusted third party, the head-end systems and the CA clients have been compromised. This is a significant advantage over all of the existing protocols, which necessitate the replacement of the entire decoder population in this scenario. The reason for this is that existing protocols require the trusted third party (or the sender) to manage a secret associated with a value that was used to initialize $B$. This secret is a key of a symmetric-key scheme or a private key of a digital signature scheme; in other words, the chip is initialized with a symmetric key or with a public key of a digital signature scheme in existing protocols. A compromise of the secret managed by the trusted third party cannot be corrected without replacing the corresponding decoder, as this is the only way to update an initialization value of $B$.

If the adversary compromises the chip of a decoder (i.e.\ compromising all its secrets), then the adversary can use the compromised chip as a source device in a control word re-distribution attack for protocol executions in which the decoder is authorized. Chips acting as source devices may be identified using a traitor tracing scheme (see e.g.\ \cite{LDR13}) and identified chips may be revoked to restore security for future protocol executions. Implicit key authentication is a useful property for the application of traitor tracing schemes as defined in \cite{LDR13} since it ensures that a compromised chip cannot be used to re-distribute control words of protocol executions in which the corresponding decoder was non-authorized. The new protocol has the property that an adversary cannot find the value of a control word if the secrets of any number of non-authorized chips are compromised. This is a useful property for identifying compromised chips in the case of a collusion attack. The adversary may also use a compromised chip as a sink device in a control word re-distribution attack. In practice, it may be hard to impossible to identify such a chip. However, for the new protocol a per-chip attack is required in order to be able to use chips as sink devices in such an attack, mitigating this threat from a practical point of view.

\section{Concluding remarks}
Section~\ref{sec:description} presented a new key establishment protocol and Section~\ref{sec:pay-TV} described how this protocol can be applied in a pay-TV system. The main innovation of the new protocol is the combination of a cryptographic hash function and a public-key encryption scheme. As shown, this results in a protocol in which the authenticity of protocol messages is protected and in which the trusted third party only needs to manage a public key of each receiver. As a consequence, the protocol achieves the unique property of enabling the restoration of security for future protocol executions in the event that the system of the trusted third party and the systems of the senders have been compromised. If applied in a pay-TV system, the new protocol reduces the risk that decoders need to be replaced to correct a security breach.

\end{document}